\documentclass[sigconf]{acmart}

\usepackage{booktabs} 
\usepackage{amsmath}
\usepackage{subfigure}
\usepackage{array}
\usepackage{balance}
\setcopyright{rightsretained}
\begin{document}
\title{Third Party Tracking in the Mobile Ecosystem}



\author{Reuben Binns, Ulrik Lyngs, Max Van Kleek, Jun Zhao, Timothy Libert\footnotemark, Nigel Shadbolt}
\affiliation{
  \institution{Department of Computer Science, University of Oxford}
  \institution{*Reuters Institute for the Study of Journalism, University of Oxford}
  \streetaddress{}
  \city{Oxford} 
  \state{} 
  \postcode{}
}
\email{reuben.binns|ulrik.lyngs|max.van.kleek|jun.zhao|nigel.shadbolt@cs.ox.ac.uk}
\email{timothy.libert@politics.ox.ac.uk}

\renewcommand{\shortauthors}{R. Binns et al.}

\begin{abstract}
Third party tracking allows companies to identify users and track their behaviour across multiple digital services. This paper presents an empirical study of the prevalence of third-party trackers on 959,000 apps from the US and UK Google Play stores. We find that most apps contain third party tracking, and the distribution of trackers is long-tailed with several highly dominant trackers accounting for a large portion of the coverage. The extent of tracking also differs between categories of apps; in particular, news apps and apps targeted at children appear to be amongst the worst in terms of the number of third party trackers associated with them. Third party tracking is also revealed to be a highly trans-national phenomenon, with many trackers operating in jurisdictions outside the EU. Based on these findings, we draw out some significant legal compliance challenges facing the tracking industry.
\end{abstract}

%

\copyrightyear{2018} 
\acmYear{2018} 
\setcopyright{acmlicensed}
\acmConference[WebSci '18]{10th ACM Conference on Web Science}{May 27--30, 2018}{Amsterdam, Netherlands}
\acmBooktitle{WebSci '18: 10th ACM Conference on Web Science, May 27--30, 2018, Amsterdam, Netherlands}
\acmPrice{15.00}
\acmDOI{10.1145/3201064.3201089}
\acmISBN{978-1-4503-5563-6/18/05}

\ccsdesc[500]{Security and privacy~Economics of security and privacy}
\ccsdesc[300]{Security and privacy~Software reverse engineering}
\ccsdesc[500]{Applied computing~Law}
\ccsdesc[300]{Networks~Mobile and wireless security}

\keywords{privacy, tracking, behavioural advertising, mobile, android, static analysis, data protection}

\maketitle

\section{Introduction}
Billions of people use smartphones every day, generating vast amounts of data about themselves. Much of the functionality afforded by these devices comes in the form of applications which derive revenue from monetising user data and displaying behaviourally targeted advertising. Firms with the ability to collect such data have become a significant part of the digital economy~\cite{acquisti2016economics}, with the online advertising industry earning \$59.6 billion per year in the U.S. alone \cite{iab2016}.

This business model is primarily enabled through `third-party' trackers ~\cite{montes2015value}, which track users via `first-party' mobile applications, whose developers embed their technology into  application source code. Such networks link activity across multiple apps to a single user, and also link to their activities on other devices or mediums like the web. This enables construction of detailed profiles about individuals, which could include inferences about shopping habits, socio-economic class or likely political opinions. These profiles can then be used for a variety of purposes, from targeted advertising to credit scoring and targeted political campaign messages.

This paper aims to provide a high-level empirical overview of the extent of third party tracking on the mobile ecosystem. In particular, we aim to answer the following:

\begin{enumerate}
\item How are third party trackers distributed across apps on the Google Play Store?\footnote{We did not study the Apple iOS App Store because there are no equivalently scalable iOS app collection and analysis methods}
\item Which companies ultimately own these tracking technologies, and in which jurisdictions are they based?
\item Do different trackers prevail amongst different genres of apps?
\end{enumerate}

Our motivation is to shed light on the status quo, in order that future efforts to address and mitigate third party tracking can be more informed and targeted. 

\section{Background}

We begin by introducing previous work on tracker detection methods, and on large-scale field studies of tracking on the web and mobile. Then, to motivate some of the present analysis, we provide an overview of existing approaches to addressing mobile tracking, including end-user controls, OS provider rules, and legal regulation. The shortcomings of the first two approaches have driven a renewed focus on the latter; by surveying the existing state of mobile tracking, we aim to provide insights into the extent to which current tracking activities may be affected by certain key data protection regulations.

\subsection{Detecting third party tracking at scale in the wild}
The third party tracking ecosystem has been studied on both the web and mobile using a variety of methods. Large scale web tracking studies detect third-party trackers by inspecting network traffic associated with a website. Some approaches use crowd-sourcing (e.g. \cite{vallina2016tracking,yu2016tracking}) while others use automated web crawlers (e.g. \cite{englehardt2016census,roesner2012detecting,libert2015exposing,yu2016tracking}. In all cases, a small number of dominant trackers are observed.

Several studies of third-party tracking have also been conducted on mobile platforms~\cite{vallina2016tracking,book2015empirical}, using both dynamic and static detection methods. Dynamic methods, as in web-based tracking studies, involve inspecting network traffic from the browser / device and identifying any third party destinations that relate to tracking. One common approach has been OS-level instrumentation, such as those of TaintDroid~\cite{enck2014taintdroid}, and AppTrace~\cite{qiu2015apptrace}.  An alternative to low-level OS instrumentation is to analyse all communications traffic transmitted by an app whilst it is in use~\cite{ren2016recon}. Other methods involve unpacking an application's source code (on Android systems, this comes as an Android Application Package (APK)) and detecting use of third-party tracking libraries~\cite{arzt2014flowdroid,batyuk2011using,egele2011pios,lin2014privacygrade}.

Other aspects of tracking have been studied, including the variety of techniques that are used, from cookies ~\cite{aziz2015cookie,englehardt2015cookies,englehardt2016census} to fingerprinting \cite{acar2013fpdetective}. A more recent field study by Yu et al. provided a finer-grained view into tracker behaviour, by classifying data being transmitted to trackers as either `safe' or `unsafe' ~\cite{yu2016tracking}. Another factor is the permissions requested by an app, which constrain the kinds of data a third party can obtain; longitudinal research has found that Android apps request additional privacy-risking permissions on average every three months ~\cite{taylor2017}.

The crossover between the mobile and web tracking ecosystem has also attracted attention in recent research. Various comparisons have shown that web and mobile tracking are different, both in terms of the companies that operate on each environment ~\cite{vallina2016tracking}, and the specific kinds of personal information that are shared by web and mobile versions of the same service ~\cite{leung2016recon}. In previous work comparing 5,000 apps and 5,000 websites, it was found that while certain companies dominate both environments, the overlap between top trackers is only partial, even for web and mobile versions of the same service \cite{binns2018measuring}.


\subsection{Existing approaches to addressing risks of tracking}
There are three main approaches for addressing the risks of tracking; end-user privacy controls, industry self-regulation, and traditional legal regulation. 

\subsubsection{End-user privacy controls}
Tracking exists on both the web and on mobile apps, but web browsers have traditionally enabled end-users to control tracking via default browser settings or through third party plugins. By contrast, no major smartphone platform OS currently gives end-users the ability to block or otherwise control third party tracking by apps (although tracker blocking is available on mobile web browsers). The privacy settings are primarily focused on app-by-app permissions, or permissions regarding certain data types (e.g. location, contacts, etc.). While various changes have been introduced like run-time permissions, and advertising identifier controls~\cite{nauman2010apex}, these do not address the distinction between first party apps and third party trackers. More recently, awareness-raising tools have been proposed which do reveal the presence of third-parties. They make use of techniques
including reverse-engineering of app source code and network traffic
analysis~\cite{batyuk2011using,qiu2015apptrace,egele2011pios,arzt2014flowdroid,enck2014taintdroid,zang2015knows,gordon2015information},
allowing identification of personal data flows from apps to first and third
parties. These tools have been used to map data flows and display them to
end-users ~\cite{balebako2013little,van2017better,srivastava2017privacyproxy,chitkara2017does}. Such
focus on third-party data collection, rather than app-level permissions, may be a more meaningful way to enact privacy choices. However, until such controls are enabled by the OS providers, third party tracking via apps remains largely invisible to end-users. This is in contrast to the web, where millions of users make use of tracker protection tools such as uBlock Origin or Ghostery.

\subsubsection{Self-regulation by platforms}

In response to the development and proliferation of trackers, and the lack of wide-scale deployment of effective end-user tracker controls, various efforts have been made by mobile OS platform developers to address the risks. Mobile application developers are required to follow the rules of the app market providers in order for their apps to be listed ~\cite{anderson2010inglorious}. Since few consumers use multiple app stores on a single smartphone, these platforms are in a stronger position to impose industry self-regulation than browser vendors, because they have the ability to effectively kick an application off the platform entirely.

Industry-led self-regulatory initiatives have thus far attempted to strike a balance between protecting users from malicious behaviour and creating a relatively permissive environment. With respect to smartphone operating systems, Apple and Google have the power to exert varying degrees of control over the behaviour of apps appearing in their default app stores. Thus far, both of their respective developer agreements permit third-party tracking, although certain user-protective practices are required, such as collecting a replaceable advertising identifier (IDFA / AAID) rather than the permanent device identifier.

More stringent action against third party tracking may also have been held back by vested interests of the OS providers. Both Google and Apple have historically had a stake in the digital advertising industry. Google own several tracker companies such as DoubleClick and others. Apple used to take a cut of advertising revenue from ad network trackers in iPhone apps, through the iADs program, but this scheme ended in 2016.




\subsubsection{Legal regulation}\label{legalbackground}

These self-regulatory efforts, such as they are, sit alongside a variety of specific legal regulations with varying levels of enforcement in different countries around the world. Perhaps the most stringent and far-sighted of these is the data protection legal regime in Europe. With updated rules incoming this year in the form of the European Union's General Data Protection Regulation, new enforcement powers including the issuing of larger fines and scope for indefinitely suspending processing may substantially curtail the activities of third party trackers.

For instance, the specific identities and purposes of third party trackers will have to be made transparent to the data subject (i.e. the user of the app); and special safeguards must be applied in the case of children. While profiling of children is not outright prohibited by the GDPR, the Article 29 Working Party (the EU body responsible for providing guidance on data protection), advise that organisations should `refrain from profiling them for marketing purposes'.

Regarding transfer of data across borders, while existing requirements are not fundamentally different under the GDPR, transnational data transfer is likely to receive additional scrutiny in light of the introduction of stronger enforcement powers. Under the existing regime, personal data is permitted to flow from one jurisdiction to another, subject to compliance with certain conditions. The least onerous condition is if the recipient organisation is based in a country whose existing data protection regime has been assessed by the European Commission and deemed `adequate'. Otherwise, special arrangements such as standard contractual clauses and binding agreements between organisations in both jurisdictions may be necessary in order to make cross-jurisdictional data flows legitimate. Similar data flow agreements exist between other countries. In some cases these are reciprocal (such as between the EU and Andorra), while others are not (e.g., the Russian privacy regulator allows personal data to flow from Russia to EU countries\footnote{\url{https://www.huntonprivacyblog.com/2017/08/16/russian-privacy-regulator-\\adds-countries-list-nations-sufficient-privacy-protections/}}, but the reverse is not true).

Such cross-border rules and data `trade blocs' have consequences for the legal basis for third party tracking when tracking companies, app developers, app stores and end-users are located in different jurisdictions. While the transfer of data from people residing in the EU to countries whose data protection regime is deemed inadequate could be legitimate in principle, more onerous conditions would need to be met. As such, any efforts to assess the legality of current practices must consider the extent to which tracking occurs across borders.



\section{Data Collection \& Methodology}

\subsection{Play Store Indexing and App Discovery}
The first step was to identify available apps. We programmatically identified popular search terms in the Play Store by autocompleting all character strings of up to a length of five, and then issued each search term to get a list of apps, ranked by popularity~\cite{google2017Autocomplete}. The identified apps were then downloaded using the \texttt{gplaycli}~\cite{gplaycli}, a command line tool for interacting with the Play Store.

\subsubsection{Static analysis method}
An Android Package Kit (APK) is an Android file format that contains all resources needed by an app to run on a device. 
Upon download, each APK was unpacked and decoded using APKTool~\cite{tumbleson2017APKTool} to obtain the app's assets, in particular its icon, bytecode (in the DEX format) and metadata (in XML format).  Finally, permission requests were parsed from the XML and hosts were found in the bytecode using a simple regex\footnote{We note that this method has the inherent problem that we cannot confirm if bytecode relating to or referencing such hosts is ever called. More sophisticated static analysis methods might better distinguish but this is left for future work. The regex used to identify hosts in the bytecode is available on \url{osf.io/4nu9e}}.


\subsubsection{Mapping hostnames to known tracker companies}

\begin{figure*}
\centering
\includegraphics[width=10cm,height=3cm]{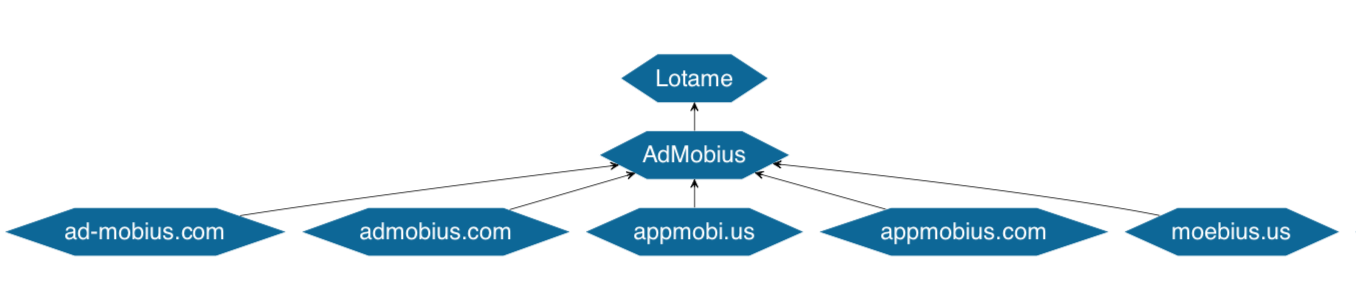}
    \caption{Example of domain-company ownership. The domain \url{Admobi.us} is owned by the company \emph{AdMobius}, which is owned by the parent company \emph{Lotame}.}~\label{fig:domain-company}
\end{figure*}

\begin{figure*}
\centering
\includegraphics[width=\textwidth,height=3cm]{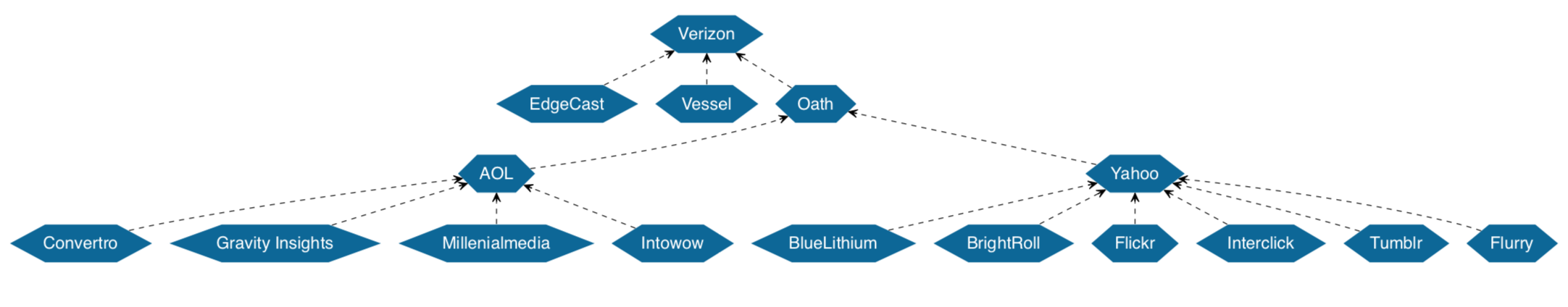}
    \caption{Example of parent-subsidiary company ownership (domains omitted). \emph{Flurry} is owned by \emph{Yahoo}, which is owned by \emph{Oath}, which is owned by \emph{Verizon} (the `root parent').}~\label{fig:parent-subsidiary}
\end{figure*}

While this static analysis process effectively identified references to hosts in the APKs, it did not provide a means of mapping them to companies, let alone selecting only those companies who are in fact engaged in tracking. A large number of the hostnames found in the static code analysis refer to a wide range of benign external resources which are not necessarily engaged in tracking. In order to isolate only those engaged in tracking, we combined two lists of trackers derived from previous research. One list is compiled by the Web X-Ray project \cite{libert2015exposing}. It maps third party web tracking domains to companies that own them, as well as parent-subsidiary relationships. The second list is compiled from previous research by the authors of the present paper \cite{binns2018measuring,VanKleek:2017:BDY:3025453.3025556}, which also maps domains to companies, and companies to their owners, but incorporates mobile app-centric trackers which are missing from web-oriented tracker lists. An example of domain-company ownership in the resulting aggregated list is shown in Figure~\ref{fig:domain-company}, and parent-subsidiary relationship in Figure~\ref{fig:parent-subsidiary}.

Host names in the tracker lists were shortened to 2-level domains using the python library \texttt{tldextract}\footnote{\url{https://github.com/john-kurkowski/tldextract}} (e.g. for `subdomain.example.com', the domain name `example' and top-level domain suffix `.com' were kept and any subdomains were omitted). Tracker hosts were then matched to hosts identified in app bytecode with a regular expression which excluded matches that was followed by a dot or an alphabetic character (matching `google.com' to `google.com/somepath' but not `google.com.domain' or `google.coming').

\subsection{Data analysis}
Most of the data analysis was conducted in R, using RStudio\footnote{Analysis scripts plus data are available via the Open Science Framework at \url{osf.io/4nu9e}. For access to the full data set, contact the authors.}.

\section{Results}
\subsection{Numbers of tracker hosts in apps}
\begin{figure}
	\centering
  \subfigure{
  \includegraphics[width=0.9\columnwidth]{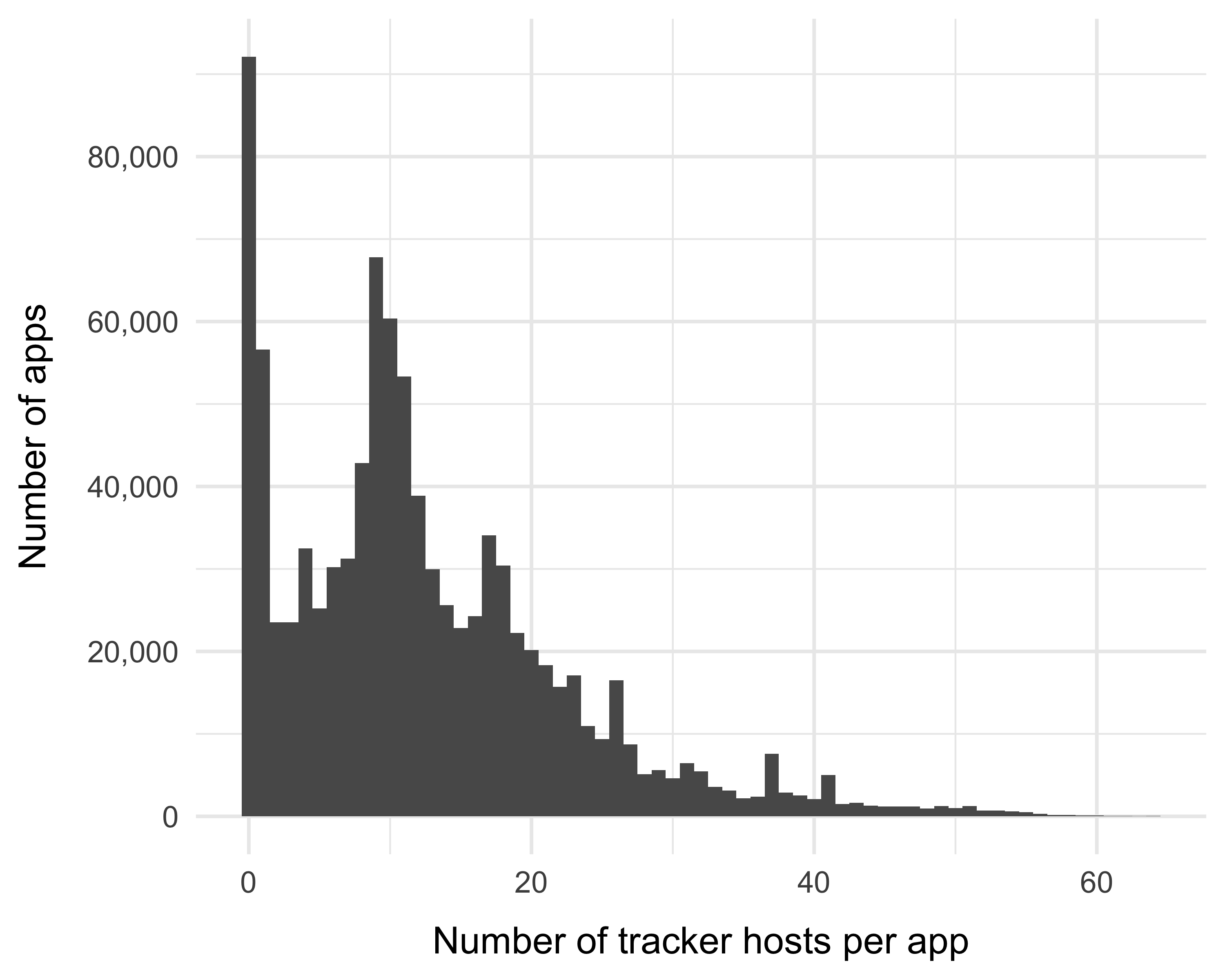}
  }
  \subfigure{
  \begin{tabular}{c c c c c}
    {\small\textit{Median}}
    & {\small \textit{Q1}}
      & {\small \textit{Q3}}
    & {\small \textit{>20 hosts}}
    	& {\small \textit{No hosts}} \\
    \midrule
    10 & 5 & 18 & 17.9\% & 9.6\% \\
  \end{tabular}
  }
  \caption{Histogram and descriptive statistics for number of tracker hosts per app (free apps on the Google Play store).}~\label{fig:trackerRefsAcrossAllApps}
\end{figure}

The distribution of number of tracker hosts per app was highly right-skewed (see Figure~\ref{fig:trackerRefsAcrossAllApps}). Gini inequality coefficient was 0.44. Across all analyzed apps (n = 959,426), the median number of tracker hosts included in the bytecode of an app was 10. 90.4\% of apps included at least one, and 17.9\% more than twenty.

\subsection{Numbers of distinct tracker companies behind hosts}
\begin{figure}
\subfigure{
\centering
  \includegraphics[width=0.9\columnwidth]{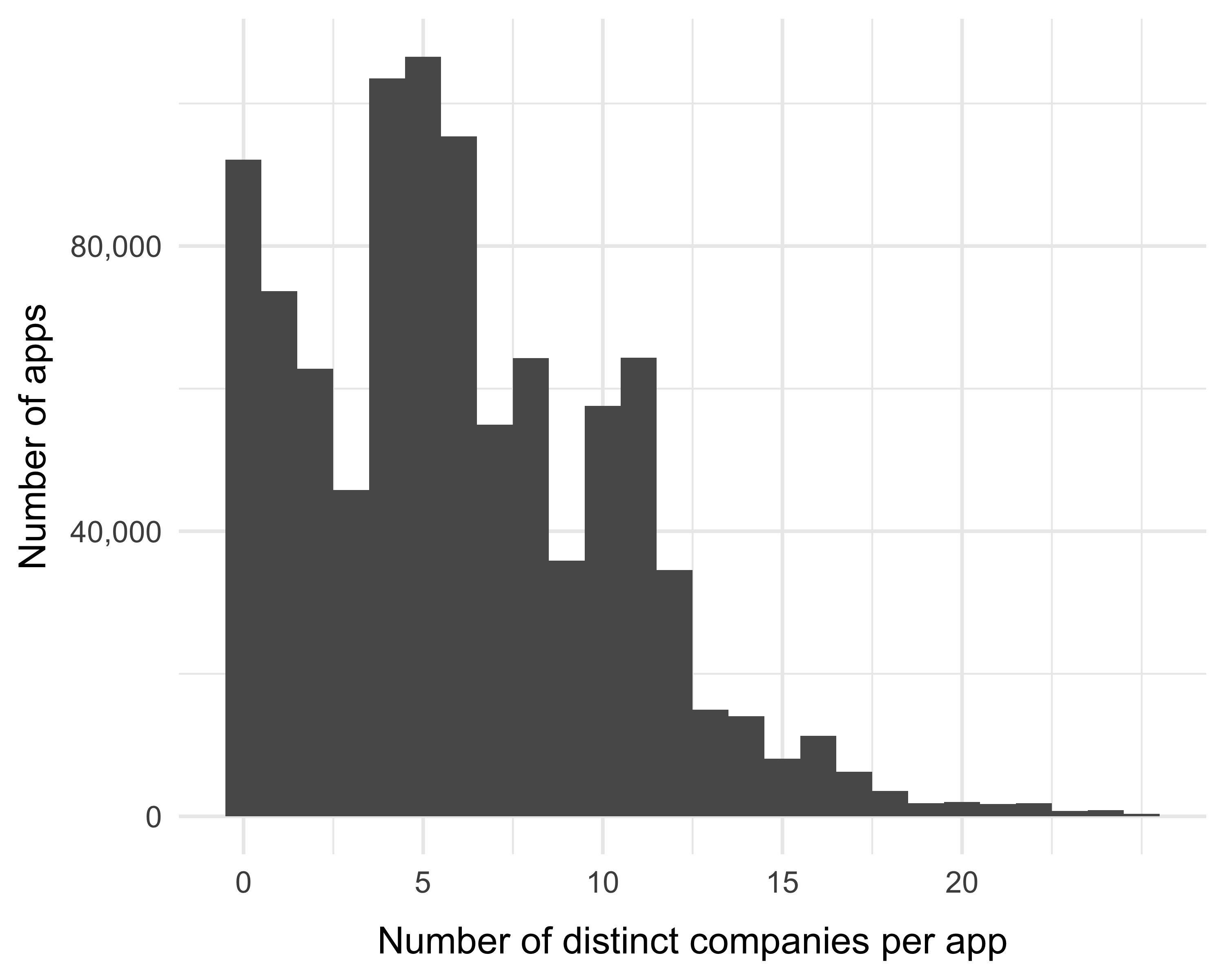}
  }
  \subfigure{
  \begin{tabular}{c c c c c}
    {\small\textit{Median}}
    & {\small \textit{Q1}}
      & {\small \textit{Q3}}
    & {\small \textit{>10 companies}}
    	& {\small \textit{No companies}} \\
    \midrule
    5 & 3 & 9 & 17.4\% & 9.6\% \\
  \end{tabular}
  }
  \caption{Number of distinct tracker companies behind hosts in apps (free apps on the Google Play store).}~\label{fig:numCompaniesReferred}
\end{figure}

The distribution of number of distinct tracker companies (at the lowest subsidiary level) behind the hosts in an app was similarly right-skewed (see Figure~\ref{fig:numCompaniesReferred}). The median number of companies was 5, 90.4\% of apps included hosts associated with at least one company, and 17.4\% with more than ten companies.

There were 13 apps for which our analysis identified 30 or more different tracking companies referred to via hosts in the bytecode. In some cases, these high numbers can be explained by the particular function of the app; for instance, some of these apps integrate multiple different services into one app (e.g. `Social Networks All in One'); in such cases, any tracking domains associated with those integrated services will be identified by our method. For others, mostly gaming apps, the high numbers of trackers serve no obvious function other than the usual kinds of behaviourally targeted advertising and analytics.

Rather than simply counting number of companies, we can query the proportion of apps containing hosts associated with specific companies. As illustrated in Figure~\ref{fig:parent-subsidiary}, however, many companies have been acquired by larger parent or holding companies, such as Alphabet. The result of grouping by 'root parent' the percentages of apps which include hosts associated with specific companies is shown in Table~\ref{tab:topCompanies}.

\begin{table}
  \begin{tabular}{>{\raggedright\arraybackslash}p{2cm} r >{\raggedright\arraybackslash}p{2cm} r c}
    {\small\textit{Root parent}}
    	& {\small \textit{\% apps}}
      & {\small \textit{Subsidiary}}
      	& {\small \textit{\% apps}}
    & {\small \textit{Country}}\\
    \midrule
Alphabet & 88.44 & Google & 87.57 & US \\ 
   &  & Google APIs & 67.51 & US \\ 
   &  & DoubleClick & 60.85 & US \\ 
   &  & Google Analytics & 39.42 & US \\ 
   &  & Google Tag Manager & 33.88 & US \\ 
   &  & Adsense & 30.12 & US \\ 
   &  & Firebase & 19.20 & US \\ 
   &  & Admob & 14.67 & US \\ 
   &  & YouTube & 9.51 & US \\ 
   &  & Blogger & 0.46 & US \\ 
  Facebook & 42.55 & Facebook & 42.54 & US \\ 
   &  & Liverail & 1.03 & US \\ 
   &  & Lifestreet & <0.01 & US \\ 
  Twitter & 33.88 & Twitter & 30.94 & US \\ 
   &  & Crashlytics & 5.10 & US \\ 
   &  & Mopub & 2.51 & US \\ 
  Verizon & 26.27 & Yahoo & 20.82 & US \\ 
   &  & Flurry & 6.28 & US \\ 
   &  & Flickr & 1.37 & US \\ 
   &  & Tumblr & 1.22 & US \\ 
   &  & Millennialmedia & 0.71 & US \\ 
   &  & Verizon & 0.11 & US \\ 
   &  & AOL & 0.06 & US \\ 
   &  & Intowow & <0.01 & US \\ 
   &  & One By AOL & <0.01 & US \\ 
   &  & Brightroll & <0.01 & US \\ 
   &  & Gravity Insights & <0.01 & US \\ 
  Microsoft & 22.75 & Microsoft & 22.11 & US \\ 
   &  & Bing & 0.12 & US \\ 
   & & LinkedIn & 20.62 & US \\ 
  Amazon & 17.91 & Amazon Web Services & 11.57 & US \\ 
   &  & Amazon & 7.72 & US \\ 
   &  & Amazon Marketing Services & 1.73 & US \\ 
   &  & Alexa & <0.01 & US \\ 
  Unitytechnologies & 5.78 & Unitytechnologies & 5.78 & US \\ 
  Chartboost & 5.45 & Chartboost & 5.45 & US \\ 
  Applovin & 3.95 & Applovin & 3.95 & US \\ 
  Cloudflare & 3.85 & Cloudflare & 3.85 & US \\ 
  Opera & 3.20 & Adcolony & 3.12 & US \\ 
   &  & Admarvel & 0.09 & US \\ 
  ~ & ~& ~&   
  \end{tabular}
  \caption{The most prevalent root parent tracking companies and their subsidiaries (full list available on \url{osf.io/4nu9e}).}~\label{tab:topCompanies}
\end{table}

\subsection{Company prevalence by genre}
The Google Play store metadata divides apps into 49 different genres (no less than 17 of these are subcategories of games, e.g. 'Casino Games' and 'Adventure Games'). To provide a high-level analysis, we grouped these genres into 8 more succinct 'super genres' (by e.g. clustering all game genres, plus the genres 'Comics', 'Entertainment', 'Sports' and 'Video Players' into a single 'Games \& Entertainment' category\footnote{See \url{osf.io/4nu9e} for details of this grouping.}). In addition, given concern of in particular tracking of children\cite{Livingstone2010}, we created a super genre consisting of apps included in one of the Google Play store's `family' categories.\footnote{All apps on the Google Play store have an ordinary genre classification, but some apps are in classified into one of the Play store's family genres.} For each super genre, we reran the company analysis, which revealed some important differences between the nature of tracking by genre.

\begin{figure}
	\centering
    \subfigure[]{
  \label{fig:descriptives}
  \begin{tabular}{l r r r r r r}
    {\small\textit{Super genre}}
    	& {\small \textit{\# apps}}
      & {\small \textit{Med.}}
      	& {\small \textit{Q1}}
    & {\small \textit{Q3}}
    	& {\small \textit{>10}}
    & {\small \textit{None}}\\
    \midrule
	News & 26281 &   7 &   4 &  11 & 29.9\% & 6.5\% \\ 
  Family & 8930 &   7 &   4 &  11 & 28.3\% & 7.2\% \\ 
  Games \& Entertainment & 291952 &   6 &   4 &  10 & 24.5\% & 7.3\% \\ 
  Art \& Photography & 27593 &   6 &   4 &  10 & 16.8\% & 3.6\% \\ 
  Music & 65099 &   6 &   4 &   8 & 13.5\% & 4.1\% \\ 
  Health \& Lifestyle & 163837 &   5 &   3 &   8 & 15.4\% & 9.0\% \\ 
  Communication \& Social & 39637 &   5 &   2 &   8 & 16.2\% & 13.4\% \\ 
  Education & 79730 &   5 &   2 &   8 & 13.3\% & 11.9\% \\ 
  Productivity \& Tools & 265297 &   5 &   2 &   8 & 11.9\% & 13.5\% \\
  \end{tabular}
  }
  \subfigure{
  \label{fig:boxplot}
  \includegraphics[width=0.9\columnwidth]{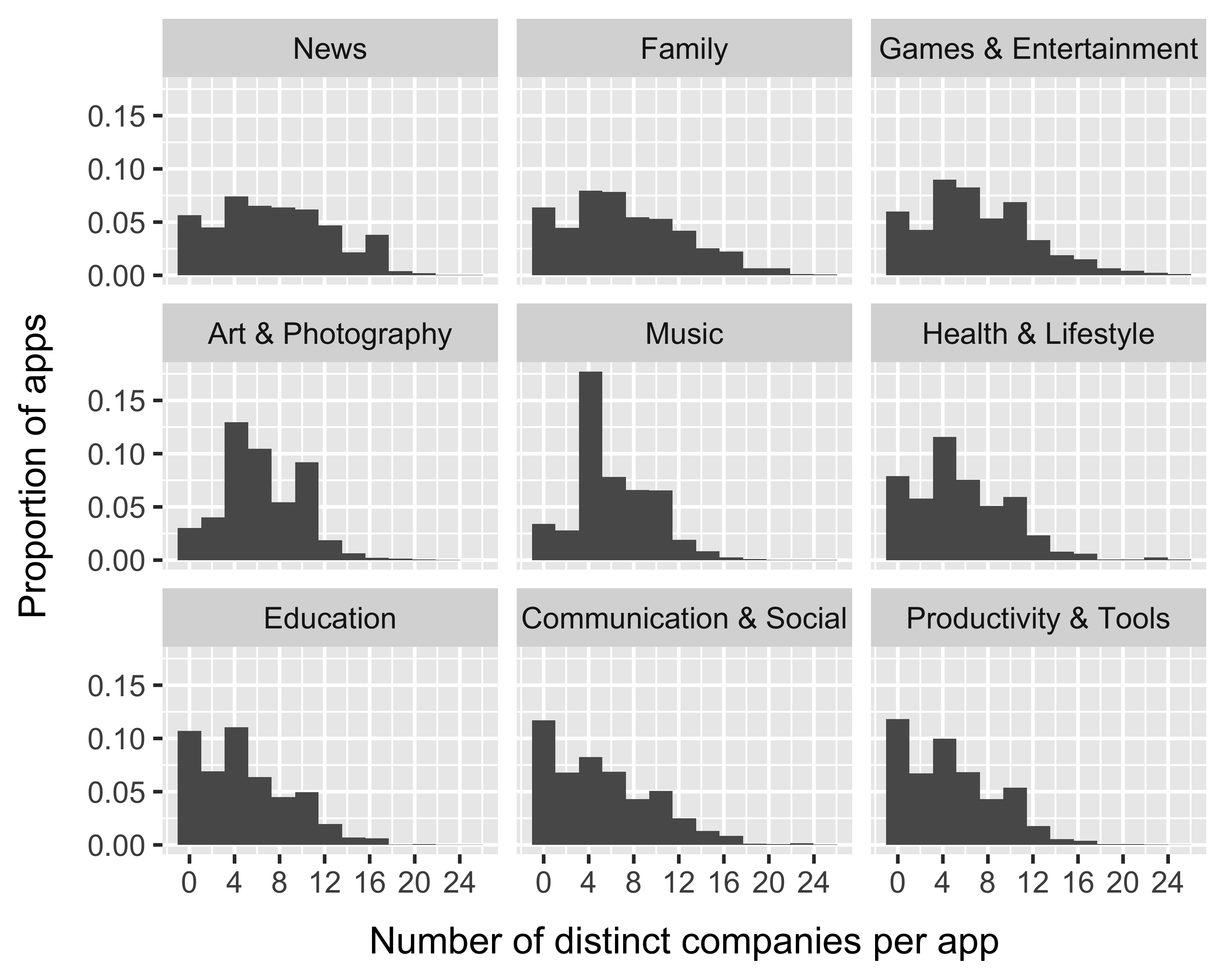}
  }
  \caption{Descriptive statistics \subref{fig:descriptives} and histograms \subref{fig:boxplot} of number of distinct tracker companies behind hosts referenced in apps, grouped by super genre.}~\label{fig:byGenreCompanyRefs}
\end{figure}

First, there are differences in the number of distinct tracking companies associated with apps from different genres. Figure~\ref{fig:byGenreCompanyRefs} shows the number of apps in each super genre, and descriptive statistics of number of distinct tracker companies associated with apps within each. \emph{News} and \emph{Family} apps have the highest median number of tracker companies associated with them, and over 20\% of apps in the \emph{News}, \emph{Family}, and \emph{Games \& Entertainment} super genres are linked to more than ten tracker companies. Meanwhile, the lowest median number of trackers are found within \emph{Productivity \& Tools}, \emph{Education}, \emph{Communication \& Social}, and \emph{Health \& Lifestyle} apps, and over 10\% of \emph{Productivity \& Tools}, \emph{Education} and \emph{Communication \& Social} apps have no trackers at all.

\begin{table}
  \begin{tabular}{p{5cm} r p{1.5cm} r p{1.5cm}}
    {\small\textit{Genre}}
    	& {\small \textit{\begin{math} K\end{math}}}
      & {\small \textit{ \begin{math}\sum{K}\end{math}}}\\
    \midrule
Productivity \& Tools & 0.14 & 5.5\\
Games \& Entertainment & 0.13 & 5.41\\
Health \& Lifestyle & 0.1 & 5.5\\
Communication \& Social & 0.09 & 5.29\\
Art \& Photography & 0.09 & 5.12\\
Family & 0.04 & 4.33\\
News & 0.03 & 4.5\\
Education & 0.03 & 5.42\\
Music & 0.02 & 5.24\\
    ~ & ~& ~& 
  \end{tabular}
  \caption[]{K distances between tracker rankings for each genre compared to all apps (K), and sum of pairwise distances between each genre and every other genre (\begin{math}\sum{K}\end{math}).}~\label{tab:genreDistances}
\end{table}

Second, there are differences in which particular trackers are associated with apps from each super genre. By comparing rankings for each, we can see the extent to which different trackers dominate each super genre. In addition to comparing the difference in rankings for any given tracker, we use an overall distance metric, the Kendall tau distance, in order to measure the extent to which rankings differ between super genres ~\cite{kendall1938new}.

The Kendall Tau distance may be defined as:
\vspace{5mm}

\begin{math}K(\tau_1,\tau_2) = \begin{matrix} \sum_{\{i,j\}\in P} \bar{K}_{i,j}(\tau_1,\tau_2) \end{matrix}\end{math}
\vspace{5mm}

where:

\begin{enumerate} 

\item''P'' is the set of unordered pairs of distinct elements in \begin{math} \tau_1 \end{math} and \begin{math}\tau_2\end{math} 
\item \begin{math}\bar{K}_{i,j}(\tau_1,\tau_2)\end{math} = 0 if ''i'' and ''j'' are in the same order in \begin{math}\tau_1\end{math} and \begin{math}\tau_2\end{math}
\item \begin{math}\bar{K}_{i,j}(\tau_1,\tau_2)\end{math} = 1 if ''i'' and ''j'' are in the opposite order in \begin{math}\tau_1\end{math} and \begin{math}\tau_2.\end{math}
\end{enumerate}

In this context, ''P'' is the set of unordered pairs of trackers (e.g. `DoubleClick' and `AdChina'),  in one genre ranking \begin{math} \tau_1 \end{math} (e.g. `Games') and another genre ranking \begin{math} \tau_2 \end{math} (e.g. `News'). \begin{math} K \end{math} is based on the number of discordant pairs between \begin{math} \tau_1 \end{math} and \begin{math} \tau_2 \end{math}, where a higher \begin{math} K \end{math} indicates greater distance.

We find that the Productivity \& Tools and Games \& Entertainment categories exhibit the biggest differences in ranking of trackers compared to the overall ranking of trackers across the whole Play Store, while the ranking of trackers in the Music category is the closest to the overall ranking (see Table \ref{tab:genreDistances}).

In addition to calculating the distance between the rankings of each genre and the rankings for the entire Play Store, we also calculated the distances between each distinct pair of genres and summed them to get an idea of the overall distance of a single genre from every other genre. When considering the distance in tracker rankings from the tracker rankings of all other categories, Productivity \& Tools and Health \& Lifestyle appear to be the biggest outliers; the top 20 trackers in the former include companies not present in the top 20 for all apps, like Mapbox (rank \#64 across all apps) as well as Chinese companies Alibaba and Baidu.

\subsection{Country differences}
\begin{figure}
	\centering
  	\includegraphics[width=0.9\columnwidth]{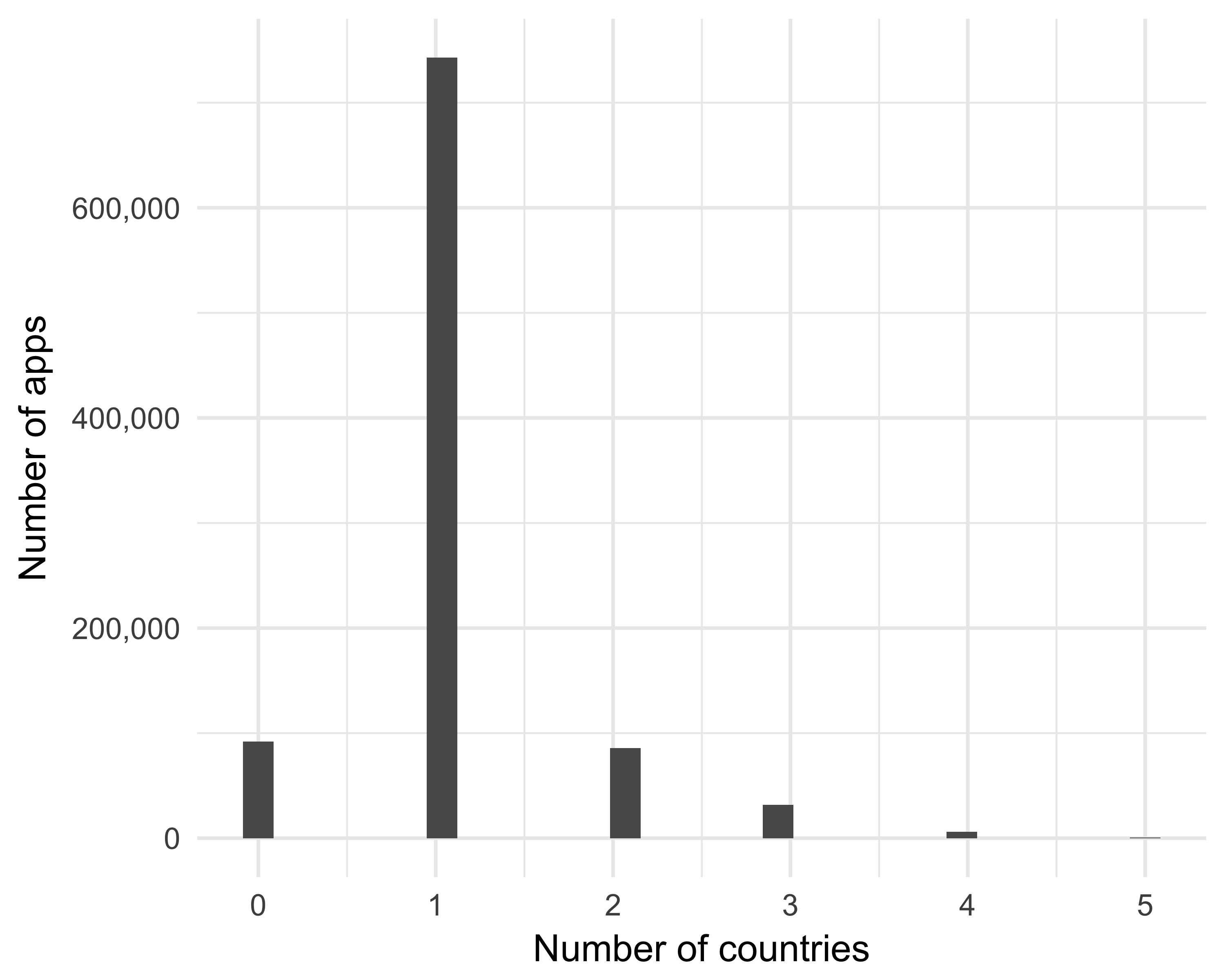}
    \caption{Number of distinct countries in which tracker companies behind hosts in an app (free apps on the Google Play store) are based.}~\label{fig:numCountriesReferred}
\end{figure}

We also analysed the prevalence of countries in which the tracker companies are based (including both subsidiary and root parent level; see Table \ref{tab:CountryPrev}). Just over 90\% of all apps contained at least one tracker owned by a company based in the United States. China, Norway, Russia, Germany, Singapore, and the United Kingdom were the next most common destinations. The median number of unique countries associated with the companies referred to in an app was 1 (see Figure~\ref{fig:numCountriesReferred}).

We also calculated the country prevalence figures on a genre-by-genre basis. While the US remained the most prevalent in every case, (between 86-96\%),  the prevalence rankings for other countries differed by super genre. For instance, UK-based trackers were the second-most prevalent in `Art \& Photography', despite being only 7th overall.

\begin{table}
  \begin{tabular}{l r r}
    {\small\textit{Country}}
    	& {\small \textit{\# apps present}}
      & {\small \textit{\% apps}}\\
    \midrule
U.S. & 865369 & 90.2\% \\ 
  China & 48451 & 5.1\% \\ 
  Norway & 30674 & 3.2\% \\ 
  Russia & 24889 & 2.6\% \\ 
  Germany & 24773 & 2.6\% \\ 
  Singapore & 19323 & 2.0\% \\ 
  UK & 14451 & 1.5\% \\ 
  Austria & 4754 & 0.5\% \\ 
  South Korea & 3366 & 0.4\% \\ 
  Japan & 1801 & 0.2\% \\
  \end{tabular}
  \caption{Apps including at least one tracker associated with a subsidiary or root parent within a given country.}~\label{tab:CountryPrev}
\end{table}

\section{Discussion}

We begin by discussing the limitations of our data collection methods. Next we consider some differences between tracking on websites and on mobile apps, and finally we draw out implications for the regulatory approaches outlined in section \ref{legalbackground}.

\subsection{Limitations of data collection methods}
There are several limitations to our tracker detection methods. First, it is incomplete; our knowledge base of tracker domain to company mappings is limited to those trackers which have been discovered in the course of previous research (namely ~\cite{VanKleek:2017:BDY:3025453.3025556,binns2018measuring,libert2015exposing}). While these lists were compiled in a systematic way, focusing on the most prevalent tracking domains, including the entire long tail of less prevalent domains might change the results reported. The inclusion and exclusion criteria for what constitutes a `tracker' are also open to debate; the list compiled in prior works, and relied on here, defines a third-party tracker as `an entity that collects data about users from first-party websites and / or apps, in order to link such data together to build a profile about the user', but the definition and its application are debateable.\footnote{The principles behind the criteria used here are discussed in the aforementioned prior works} Another issue is that without dynamic network traffic analysis of all apps, including successful man-in-the-middle proxying and ability to interpret the data payloads, we cannot confirm precisely what data is sent to each tracker. Finally, different trackers serve different purposes; some facilitate targeted advertising, while others are used for analytics. Without further fine-grained distinctions between such purposes, the figures presented here do not represent the full nuance and variety of third party tracking and its impacts.

\subsection{Web vs. Mobile}
Previous large-scale studies of tracking have largely focused on the web. The distribution model of the web allows measurement of tracking to scale in a way that the model for smartphone app distribution does not; web services are delivered in a standardised way through a browser which can easily be automated. As a result, large-scale web tracking studies typically include millions of sites. By contrast, the largest smartphone app tracking study to our knowledge at the time of writing is derived from network traffic detected by the Lumen app, which includes the data flows of 14,599 apps installed on Lumen user's devices \cite{razaghpanah2018apps}. While such crowdsourced methods have many advantages in terms of the granularity of the data flows and ecological validity, at best they scale to tens of thousands of apps. By contrast, our method is scalable to hundreds of thousands of apps (indeed, our dataset of apps is close to a million).

\subsection{Implications for tracker regulation}

While the distribution of trackers across apps is of general interest from a privacy and data protection regulation perspective, we focus here on several particular regulatory implications arising from our findings.

\subsubsection{Cross-jurisdictional data flow} As explained in Section \ref{legalbackground}, the rules regarding transfers of data outside the EU under the GDPR are similar to the previous regime (under the Data Protection Directive), but with some new details as well as larger associated fines. In so far as these developments result in more investigation and enforcement by authorities, the impact will be different for companies depending on their jurisdiction. There will be no impact on those based in the EU, such as Germany (the fifth-most prevalent country in which trackers are based), who benefit from rules permitting the free flow of data within the Union. Some third countries such as Canada also benefit from being on the EU Commission's list of legal regimes that are deemed `adequate' and therefore data transfers to trackers in those jurisdictions are legitimate without further measures in place.

However, amongst the top-10 most prevalent countries there are several which lie outside the E.U. and are not deemed adequate, such as China, Russia, Singapore, South Korea and Japan. In order for transfers to these countries to be legitimate, additional safeguards must be in place as explained in Section \ref{legalbackground}. We cannot determine whether such arrangements have been put in place by the identified companies based in non-approved jurisdictions, but our figures give an indication of the volume of companies to whom these more onerous rules apply. While the percentages of apps which include trackers from such jurisdictions are small compared to the US---China (5.1\%), Russia (2.6\%), Singapore (2\%) versus US (90\%)---they are still significant, numbering in the tens of thousands. 

\subsubsection{Profiling}
The GDPR uses the term `profiling' to describe any fully or partly automated processing of personal data with the objective of evaluating personal aspects of a natural person (Article 4(4)). Many of the tracking companies included in our knowledge base engage in data processing activity that would likely constitute `profiling' under this definition. For instance, the purpose of many of the most common trackers is behaviourally targeted advertising, whereby individuals are evaluated along demographic and behavioural dimensions to determine their propensity to respond to certain marketing messages. Profiling is prohibited if it has `legal or significant' effects on the data subject. While the definition of `significant effects' is not entirely clear, the Article 29 Working Party has advised that even profiling for marketing purposes could potentially give rise to significant effects, including if it is: intrusive; targets vulnerable, minority groups, or those in financial difficulty; involves differential pricing; or deprives certain groups of opportunities.\footnote{Article 29 Working Party: Guidelines on Automated individual decision-making and Profiling for the purposes of Regulation 2016/679 \url{http://ec.europa.eu/newsroom/article29/item-detail.cfm?item_id=612053}} Trackers which enable such activities without consent of the data subject could therefore be in breach of Article 22 (unless such profiling is necessary for entering or performing a contract, or it is authorised by another member state law). Many of the most prevalent trackers observed in our study have the capacity to be used in such ways, and evidence of such practices is beginning to emerge. For instance, DoubleClick (present on 60\% of apps analysed) has been shown to target adverts for higher-paid jobs to men at a higher rate than to women \cite{datta2015automated}; while web-based price discrimination has also been documented by numerous studies in recent years \cite{mikians2012detecting,hannak2014measuring}.

\subsubsection{Rights and obligations regarding children}
Like the old Directive, the GDPR defines certain additional rights and obligations regarding processing the personal data of children (defined as anyone under the age of 16, and for certain additional protections, 13). If a tracker is relying on consent as a legitimating ground for processing, then such consent would not be valid from a child under 13; instead a parent or guardian would need to consent. Furthermore, as discussed above, Recital 38 states that special protections should be in place if children's data are being processed for marketing and user profiling. This description would likely cover many of the trackers which are embedded in apps from the Family and Games \& Entertainment genre categories, which are clearly targeted at children. Problematically, apps from these two genres are especially exposed to third party tracking, with the average app including hosts associated with 7 distinct tracker companies for Family apps, and 6 for Games \& Entertainment apps (only News apps are more exposed). Given the relatively higher level of protection set in the law regarding profiling children for marketing, it seems that tracking is most rampant in the very context in which regulators are most concerned to constrain it.

\section{Conclusion}
We believe that by undertaking analysis of the distribution of tracking technology on close to 1 million smartphone apps, we gain insight into the breadth and scale of this highly important phenomenon. Unlike previous studies whose coverage of apps numbers in the tens of thousands, and may be skewed towards the app choices of the users from whom data is gathered, our study is a systematic analysis of apps on the Play Store.

Our genre-by-genre analysis suggests that there are differences in the behaviour and distribution of trackers depending on the functionality or purpose the app provides. News and Games apps appear amongst the worst in terms of the number of tracker companies associated with them. Tracking is also a substantially trans-national phenomenon; around 100,000 apps we analysed send data to trackers located in more than one jurisdiction. 

These findings suggests that there are challenges ahead both for regulators aiming to enforce the law, and for companies who intend to comply with it. Full audits of mobile app stores such as this could help regulators identify areas to focus on. Previous privacy enforcement `sweeps'\footnote{See \url{https://www.privacyenforcement.net/node/906
}} have focused on the most popular apps, and their terms of service and privacy policies. But the analysis here suggests that apps may not necessarily be the most efficient point of analysis; rather, identifying and investigating the most prevalent trackers might be a better target. Some of the practices likely to be involved - such as allowing profiling of children without attempting to obtain parental consent - may be downright unlawful. It remains to be seen how and if regulators will attempt to detect and prevent behavioural targeting that has `significant effects' on data subjects.

The governance of these activities is complex, involving many stakeholders, including: users, smartphone operating system developers, equipment manufacturers, alternative app market operators, app developers, and tracking companies (who also operate multi-sided markets with advertisers and therefore have the ability to impose constraints on what ads can be served). Effective regulation will require collaboration between regulators and these myriad other actors.

\begin{acks}
All authors are supported under \emph{SOCIAM: The Theory and Practice of Social Machines}, funded by the UK Engineering and Physical Sciences Research Council (EPSRC) under grant number EP/J017728/2 and comprises the University of Oxford, the University of Southampton, and the University of Edinburgh. Reuben Binns and Max Van Kleek are also supported by \emph{ReTiPS: Repectful Things in Private Spaces}, a project funded through the PETRAS IoT Hub Strategic Fund, which, in turn, was funded by the EPSRC under grant number N02334X/1. Timothy Libert is also supported by the Google Digital News Project at the Reuters Institute for the Study of Journalism. Jun Zhao is also supported by KOALA (http://SOCIAM.org/project/koala): Kids Online Anonymity \& Lifelong Autonomy, funded by EPSRC Impact Acceleration Account Award, under the grant number of EP/R511742/1.
\end{acks}

\bibliographystyle{ACM-Reference-Format}
\balance
\bibliography{bibliography} 

\end{document}